\documentclass[11pt,oneside,letterpaper]{article}
\usepackage{amssymb}
\usepackage{amsmath,stackengine}
\usepackage[dvips]{graphicx}
\usepackage{setspace}
\usepackage{amsfonts}
\usepackage{fancyhdr}
\usepackage{xcolor}
\usepackage{mathrsfs}
\usepackage{graphicx}
\usepackage{rotating}
\usepackage{comment}
\usepackage{fancyvrb}
\usepackage{color}
\usepackage{cite}
\usepackage{braket}
\usepackage{caption}
\usepackage[utf8]{inputenc}
\usepackage[c]{esvect}
\usepackage{mathabx}
\usepackage{framed}
\usepackage{scalerel}
\usepackage{tikz}
\usepackage{tcolorbox}
\usepackage[hidelinks,colorlinks=true,linkcolor=blue,citecolor=blue]{hyperref}
\usepackage{textcomp}

\tcbuselibrary{breakable}

\usepackage{mwe}

\definecolor{lightgray}{gray}{0.85}

\definecolor{darkgreen}{rgb}{0,0.5,0}
\definecolor{darkblue}{rgb}{0,0,0.6}
\definecolor{purple}{rgb}{0.4,.2,0.7}

\newcommand{\be}{\begin{equation}}
	\newcommand{\ee}{\end{equation}}

\newcommand{\chiTwo}{\text{\raisebox{0.3ex}{\scalebox{0.85}{$\chi$}}}}

\usepackage{pdfsync}
\makeatletter
\newcommand*{\mybox}{\mathrel{%
			\raisebox{0.25ex}{$\Box$}}%
	} 
\makeatother
\DeclareMathOperator{\diag}{diag}

\def\be{\begin{eqnarray}}
	\def\ee{\end{eqnarray}}

\newcommand{\bea}{\begin{eqnarray}}
	\newcommand{\eea}{\end{eqnarray}}
\def\ben{\begin{equation}}
	\def\een{\end{equation}}

     \let\r=v

\def\be{\begin{equation}}
	\def\ee{\end{equation}}
\def\ba{\begin{array}}
	\def\ea{\end{array}}

\def\ba#1\ea{\begin{align}#1\end{align}}
\def\bs#1\es{\begin{split}#1\end{split}}

\allowdisplaybreaks  % allow page breaks in displayed eqs

\numberwithin{equation}{section}
\def \be {\begin{equation}}
	\def \ee {\end{equation}}
	
\def \JM#1 {{\color{blue}  JM: #1 }}
\def \AAl#1 {{\color{red}  AA: #1 }}

\begin{document}
	\onehalfspacing
	
	\begin{center}
		
		~
		\vskip5mm
		
		{\LARGE  {Cosmological perturbations meet Wheeler DeWitt
		}}
			\vskip10mm
			 Federico Piazza and  Siméon Vareilles

		{\it Aix Marseille Univ, Universit\'{e} de Toulon, CNRS, CPT, Marseille, France}
			%} 
		
		\vskip5mm

		%{\tt piazza@cpt.univ-mrs.fr }
		
		\href{mailto:piazza@cpt.univ-mrs.fr}{\tt piazza@cpt.univ-mrs.fr}, \ \ \ \ \href{mailto:vareilles@cpt.univ-mrs.fr}{\tt vareilles@cpt.univ-mrs.fr}
	\end{center}

	\vspace{4mm}
	
	\begin{abstract}

 We study approximate solutions  of the Wheeler DeWitt (WdW) equation and compare them with the standard results of cosmological perturbation theory. In mini-superspace, we introduce a dimensionless gravitational coupling $\alpha$ that  is typically very small and functions like $\hbar$ in a WKB expansion. We seek solutions of the form $\Psi = e^{iS/\alpha} \psi$  that are the closest quantum analog of a given classical background spacetime. 
	The function $S$ satisfies the Hamilton-Jacobi equation, while $\psi$ obeys a Schr\"odinger-like equation and can be given  a probabilistic interpretation. The semiclassical limit suggests a specific relation between $\psi$ and the standard perturbation-theory wavefunction $\psi_P$. We verify this relation in two main  examples: a scalar field with a purely exponential potential, of which simple scaling solutions are known and a slow-roll scenario expanded in the vicinity of the origin in field space. Each example is worked out in two different gauges, that are the minisuperspace equivalent of unitary gauge and spatially flat gauge.  We discuss possible deviations from the classical background trajectory as well as the higher ``time" derivative terms that are present in the WdW equation but not in the perturbative approach. We clarify the \emph{conditional probability} content of the wavefunctions and how this is related with the standard gauge fixing procedure in perturbation theory.

 	 \end{abstract}
%\vspace{.2in}
%\vspace{.3in}

\pagebreak
\pagestyle{plain}

\setcounter{tocdepth}{2}
{}
\vfill

\ \vspace{-2cm}
\newcommand{\deq}{{\overrightarrow {\Delta x}}^{\, 2}}
\renewcommand{\baselinestretch}{1}\small
\tableofcontents
\renewcommand{\baselinestretch}{1.15}\normalsize

%\section{What state is the state of the universe in?}

\section{Introduction}

It is common practice in cosmology to study the behavior of small perturbations around a classical background solution of the Einstein equations. These perturbations can and should be considered at the full quantum level, as suggested by the standard inflationary paradigm. 
The purpose of this paper is to better---and somewhat pedantically---understand the relation between cosmological perturbation theory and 
 the Wheeler DeWitt (WdW) equation~\cite{DeWitt:1967yk}. The latter is widely believed to govern the wavefunction of the universe non-perturbatively, \emph{i.e.} without any reference to a classical background, at least in the sub-Plankian regime of validity of general relativity.  There is little doubt that perturbative quantum gravity is contained in the WdW equation in some way, but showing this relation explicitly is not necessarily straightforward. In Minkoswki and AdS spaces this has been done in~\cite{Kuchar:1970mu} and~\cite{Chakraborty:2023yed} respectively.  Cosmological spacetimes differ from Minkowski and AdS in at least two regards.

First, in cosmology there is one or more dominant components (e.g. radiation) driving the expansion. We cannot take the semiclassical limit $G\!\rightarrow\! 0$ of the WdW equation too naively, or we would end up  decoupling matter from gravity and expanding around a \emph{vacuum} solution of the Einstein equations. This is the wrong saddle point for cosmology. The correct semiclassical limit is discussed below in Sec.~\ref{minisuper}. 

Related to this, as a consequence of the spontaneously broken time translations~\cite{Creminelli:2006xe,Cheung:2007st,Piazza:2013coa} cosmology is the realm of the gravitational adiabatic scalar mode, which is not present in empty Minkowski and AdS spaces. The mini-superspace approximation adopted in this paper eliminates the on-shell gravitons, i.e. the tensor modes, but keep some version of the scalar mode alive.

Secondly, cosmological spacetimes are time dependent. In general, quantum  averages evolve in time differently than classical solutions, 
so one could end up recovering perturbative quantum gravity around a (sightly) different background! This is a complication  and, at the same time, a potential insight of the non-perturbative WdW approach.  We expand on this aspect in Sec.~\ref{deviat} below. 

The key objects to compare between the two approaches are their respective wavefunctions and the equations that govern them. In fact, while equal-time correlators remain the most extensively studied object in cosmology, there is growing interest in exploring the \emph{perturbative wavefunction of the universe} $\psi_P$ and its fascinating analytical properties (e.g.~\cite{Salcedo:2022aal,Benincasa:2022gtd} and references therein). This can be considered as the most primitive of all cosmological observables, the one from which all correlators can be derived. It satisfies a Schr\"odinger equation 
\begin{equation} \label{schintro}
i \partial_t \psi_P = H \psi_P\, . 
\end{equation}
In the above, $\psi_P$ is a functional of the perturbation variables, which on a cosmological background can be split into scalars, vectors and tensors according to their helicities.  The Hamiltonian $H$ should be worked out at the required order of approximation in these perturbations. We would like to understand the relation between $\psi_P$ and  the WdW wavefunction $\Psi$, satisfying 
\begin{equation}
{\cal H} \Psi = 0\, , 
\end{equation}
where $\cal H$ is the full Hamiltonian of the system (e.g. eq.~\eqref{hamiltonianwdw} below).

One main limitation of this paper is the use throughout of the mini-superspace approximation. It seems to us that all subtleties that are peculiar of dealing with a cosmological setup are basically already present at the mini-superspace level. Extending our result to the ``real thing" should be relatively straightforward, at least conceptually.  
At the same time, despite its obvious limitations (see e.g.~\cite{Nicolis:2022gzh}), the mini-superspace approach should have its own interesting regime of applicability, as argued in Sec.~\ref{minisuper}.

The paper is organized as follows. In Section~\ref{sec_preliminary}, we review the minisuperspace approximation and some aspects of standard perturbation theory.  In Section~\ref{sec_WdW_ViewPoint}, we turn to the WdW equation. We discuss the probabilistic interpretation of the associated wavefunction. A relation between the latter and the wavefunction of perturbation theory $\psi_P$ is proposed and several differences between the two approaches analysed. 
 These differences are then highlighted in detail in the case of a scalar field. In Sec.~\ref{dom_component} we describe our two scalar field models and their standard (albeit mini-superspace-) treatment in perturbation theory. Then we approach these models with the WdW equation.  The first model (Sec. \ref{scaling}) is a purely exponential potential that allows a complete analytic study of the WdW equation. The second model is a slow-roll potential  (Sec.~\ref{slowroll}), that we work out in the slow roll approximation in the vicinity of the origin in field space.  We finally draw some conclusions in Sec.~\ref{conclusion}.

\subsubsection*{Note on Bibliography}
Much of this paper builds on ideas and techniques that are scattered through the extensive literature on the WdW equation.  We have learned a great deal  from Refs.\cite{Halliwell:1988wc,Vilenkin:1988yd,Halliwell:1990uy,Janssen:2020pii,Hartle:1992as,maldavideo,Lehners:2023yrj,Maldacena:2024uhs} (many of these works focus on the no-boundary proposal~\cite{Hawking:1981gb,Hartle:1983ai}, while here in we are mostly concerned with the \emph{evolution} of the state of the universe, not with its beginning). The content of this paper also clearly overlaps with~\cite{Brizuela:2016gnz,DiGioia:2019nti,Chataignier:2019psm,Kamenshchik:2020yvs,Chataignier:2020fap,Chataignier:2022iwb,Maniccia:2023cgv,Chataignier:2023rkq}. We leave to future work a more detailed comparison between the present approach and such references.

\section{Preliminary considerations }\label{sec_preliminary}

As preliminary to our analysis,  we find it helpful to spell out some of the scales that are often set conventionally to one in mini-superspace, and discuss the limits of applicability of such an approximation. 
%At the same time, we are going to argue for  to set the perimeter of applicability 
%In particular, $H_\star$ is a reference Hubble parameter, $\ell \sim M_P^2 H_\star^{-3}$ a (very super-Hubble) length and $\alpha \equiv  1/(\ell H_\star)$ the dimensionless coupling of the quantum theory.  ($M_P = 1/\sqrt{8 \pi G}$ is the reduced Planck mass.) 

\subsection{(Super-Hubble) patches as physical systems} \label{minisuper}

The mini-superspace approximation best applies to a spatially closed universe, where the scale factor $a$ can be  identified with the ``zero-mode" of the system---the total size of the spatial slices. However, the same formalism should also be able to capture the average expansion of a \emph{patch} of universe, if the patch is sufficiently homogeneous and isotropic. 
 The latter properties seem to characterize the super-Hubble regions of our universe at any time.\footnote{These considerations are reminiscent of, and ideally supported by, the \emph{separate-universe} approach in cosmological perturbation theory (e.g.~\cite{Salopek:1990jq,Sasaki:1995aw,Wands:2000dp,Pattison:2019hef,Artigas:2021zdk}).}
 
 Say we want to describe the system at around some time $t_\star$ when we can set the scale factor to $a = e^\rho =1 $ (we are taking the spatially flat limit where $ds^2 = - N^2 dt^2 + e^{2\rho}d\vec x^2$). We can then consider  a region of about the  Hubble size $\sim H_\star^{-1}$ and follow its comoving evolution as long as it stays outside the horizon. For an accelerating universe, this corresponds to following the evolution \emph{forward in time}, as comoving regions tend to fall out of causal contact.  In a decelerating universe we can track the \emph{backward} evolution of the comoving region. 

By integrating the Einstein Hilbert action over an approximately homogeneous, comoving spatial volume of size $\gtrsim H_\star^{-3}$ one gets an action for $\rho$,\begin{equation} \label{scalar_intro} 
\frac{M_P^2}{2}\int d^4x\,  \sqrt{-g}  R + boundary \  terms \ \simeq \ L \int dt \, e^{3 \rho}\left(- \frac{3 \dot \rho^2}{N}\right)
\end{equation}
%\begin{equation}
%    I=\frac{1}{\alpha H_{\star}} \int dt \, e^{3\rho}\bigg(-\frac{3\dot{\rho}^2}{N}+\frac{\dot{\phi}^2}{2N}-NH_\star^2V(\phi)\bigg).  \label{scalar_intro} 
%\end{equation}
where $M_P = 1/\sqrt{8 \pi G}$ is the reduced Planck mass and $L\sim M_P^2 H_\star^{-3}$ is a (very super-Hubble) length. %The spatial integration of the original action over at least a Hubble volume has produced the very large length $\frac{1}{\alpha H_{\star}} \sim M^2_{P} H_\star^{-3} $. .
With the aim of including matter fields in~\eqref{scalar_intro}, it is useful to consider directly a more general model (e.g.~\cite{Janssen:2020pii}),
\begin{equation} \label{action_general}
I = L \int dt \, \left(\frac{1 }{2 N} g_{\mu \nu} \dot q^\mu \dot q^\nu - N H_\star^2 U(q^\mu)\right),
\end{equation} 
where $q^\mu$ are $d+1$ (dimensionless) dynamical variables and the potential $U$ is normalized to be of order one at  $t_\star$, so that the classical Friedman equation  yields the Hubble parameter $\dot \rho(t_\star) \simeq H_\star$. 
The metric $g_{\mu \nu}$ has signature $(-,+,\dots, +)$ because the scale factor $\rho$ (represented here by $q^0$) appears in the action with the wrong sign.  We highlight that, here and  throughout the paper, $g_{\mu\nu}$ is the \emph{field-space metric}, which has nothing to do with the spacetime metric. 

The classical solutions  satisfy the constraint equation
\begin{equation} \label{constraineq}
g_{\mu \nu} \, \dot {q}^\mu \dot { q}^\nu + 2 H_\star^2 U\, N^2= 0\, ,
\end{equation}
obtained by varying~\eqref{action_general} with respect to $N$. As apparent from the form of the spacetime metric, $N=1$ corresponds to choosing the standard proper time of cosmological observers as a clock, although other choices are also possible (see e.g. in the subsection below). When $N=1$, the equations of motion read
\begin{equation} \label{fieldeq}
\ddot q^{\, \mu} + \Gamma_{\nu \sigma}^\mu \,  \dot q^\nu \dot q ^\sigma = - H_\star^2 \, \partial^\mu U  \, .
\end{equation}
Were it not for the potential, this is just the geodesic equation in the minisuperspace $g_{\mu \nu}$ metric.
We indicate with $\bar q^{ \, \mu }(t)$ a solution of~\eqref{constraineq}-\eqref{fieldeq} and, more specifically, \emph{the} solution that we want to describe quantum mechanically in the WdW formalism.

As mentioned in the introduction, when we take the \emph{semi-classical limit} in cosmology, we want to make sure that the fields keep feeding the metric while $M_P\rightarrow \infty$. With the fields $q^i$ already defined to be dimensionless, this is simply achieved by keeping $H_\star$ constant in the process, as clear from~\eqref{constraineq}.  

In the WdW equation, $L$ and $H_\star$ only enter in the dimensionless combination 
\begin{equation} \label{alphaa}
\alpha \ \equiv \  \frac{1}{L H_\star} \sim \frac{H_\star^2}{M_P^2}\, .
\end{equation}
This is the ``loop coupling" parameter taking the place of $\hbar$ in the calculations. In the rest of the paper, $L$ will be dropped in favor of $\alpha$. The semiclassical limit corresponds to 
\emph{sending $\alpha$ to zero while keeping $H_\star$ constant}.
 
 Notice that the value of $\alpha$ depends on the system. If we take $t_\star$ during the observable window of inflation, from the observed primordial power spectrum one can estimate that $\alpha \sim 10^{-9}  \epsilon$, with $\epsilon$ the slow-roll parameter.  Regions exiting during eternal inflation can have a much higher $\alpha$, depending on the model. In fact, the condition for eternal inflation can be rephrased as $\alpha/\epsilon \gtrsim1$. For a quadratic potential one can estimate $\alpha \gtrsim 10^{-5}$, which reiterates that the system-universe is still perturbative during this phase~\cite{Creminelli:2008es}. 
The smallness of $\alpha$ after inflation shows the unbearable irrelevance of late-time quantum gravity effects. 

\subsection{Perturbation theory in mini-superspace }

Despite the presence of $d+1$ fields $q^\mu$, the system~\eqref{action_general} has only $d$ degrees of freedom. One way to see this  is to notice that the constraint~\eqref{constraineq} can be used to find $N$, 
\begin{equation} \label{N_2}
N = \pm \, \sqrt{ \frac{- g_{\mu \nu} \, \dot {q}^\mu \dot { q}^\nu}{ 2 H_\star^2 U}}\, .
\end{equation}
By inserting this expression back in action we obtain
\begin{equation}
I = - \frac{\sqrt{2}}{\alpha} \int dt \, \sqrt{ - g_{\mu \nu} \, \dot {q}^\mu \dot { q}^\nu \,  U}\label{intro_squareroot}\, ,
\end{equation}
where the $+$ branch has been chosen in~\eqref{N_2}. 
In the last equation, the kinetic terms appear inside a square root, making explicit that the theory is invariant under time-reparameterizations $t \rightarrow \tilde t(t)$. In particular, time can be directly identified with one of the fields, say $ t = q^0 H_\star^{-1}$. This is a possible \emph{gauge fixing} in this mini-superspace framework.  One is then left with only the $d$ ``spatial" degrees of freedom $q^i$, governed by the action  

\begin{equation} 
I = - \frac{\sqrt{2}}{\alpha} \int dq^0 \, \sqrt{ - \left(g_{00} + 2 g_{0i} \partial_0 q^i + g_{ij} \partial_0 q^i \partial_0 q^j\right) \,  U}\label{intro_squareroot2}\, .
\end{equation}

The perturbations around a classical background solution  of~\eqref{intro_squareroot2}, $\bar q^{\, i}(q^0)$, can be quantized straightforwardly. Notice that, in a full cosmological setup, one choses the background on symmetry principles (homogeneity  and isotropy). At this mini-superspace level, the distinction between background and perturbations is completely artificial, so $\bar q^{\, i}(q^0)$ is simply \emph{one} possible solution.  By defining 
\begin{equation}\label{varphis}
\varphi^i = q^i - \bar q^{\, i}(q^0)\, ,
\end{equation}
one can expand~\eqref{intro_squareroot2} to quadratic order in $\varphi^i$. The quantization proceeds in the standard way by promoting the canonical variables $\varphi^i$  to operators and identifying the conjugate momenta with $p_i \sim - i \partial_{\varphi^i}$.

The perturbation theory wavefunction $\psi_P$  evolves in time with the quadratic Hamiltonian and thus obeys a Schr\"odinger equation of the type
\begin{equation} \label{sch_sec2}
i  \, \frac{\partial \psi_P}{\partial q^0} = - \frac{\alpha}{2} \, \nabla^2_{\varphi^i} \, \psi_P + \frac{1}{\alpha}\frac{M_{ij}(q^0) \varphi^i \varphi^j}{2} \, \psi_P\, , 
\end{equation}
with $\nabla^2_{\varphi^i}$ a suitably defined ``spatial" Laplacian that depends on $q^0$ and $M_{ij}$ a mass term.

This is the type of equation  that we would like to recover within the WdW formalism. It is already written in a ``relational" form, in the sense that one of the variables is used as time. There is a very cheap way to recover standard perturbation theory in cosmic time $t$ (\emph{e.g.} eq.~\eqref{schintro}), which is making use of the background solution $\bar q^{\, \mu}(t)$ in order to express $q^0$ in terms of $t$.  We are still in a gauge in which $q^0$ is not fluctuating, so this corresponds to fixing $N = 1+ \delta N$, with $\delta N$ such that $\delta q^0 = 0$ (see the \emph{unitary} ($\delta \phi = 0$) and \emph{spatially flat} ($\delta \rho = 0$) examples throughout the paper).

This way of introducing the proper time of the cosmological observers presumes a classical one-to-one correspondence between $q^0$ and $t$. In a fully-quantum cosmology setup time should instead be introduced as an additional degree of freedom (e.g.~\cite{Goeller:2022rsx,Witten:2023qsv,Geng:2024dbl} for recent discussions). This will be done in a subsequent paper. 

\subsection{Gauge fixing and conditional probability}

To the perturbative wavefunction $\psi_P$ is associated the standard probability density
\begin{equation}\label{probapp_intro}
d P(  q^i \, |q^0)  =  \ |\psi_P|^2 \, d^d q\, .
\end{equation}
In fact, the Schr\"odinger wavefunction $\psi_P$ is naturally interpreted in terms of \emph{conditional} probability---to find the system in some configuration $q^i$ \emph{given that} the clock measures $q^0$. However, already in perturbation theory, what we decide to call ``time" is a mere gauge choice. At the level of the wavefunction fixing the gauge can really be seen as deciding which combination of variables to use as a ``condition" in the conditional probability expression. This is even more striking from the perspective of the WdW equation (see also~\cite{Piazza:2021ojr} on this), in the sense that $\Psi(q^\mu)$ is completely agnostic about gauge choices. However, in order to extract a probability from it, one needs to decide how to foliate the field space, or, in other words, which combination of coordinates should be used as ``time" (see Sec.~\ref{probabilities}). 

For the scalar field example used in this paper (see action~\eqref{eq:action_model1} below) the situation is depicted in Fig.~\ref{fig_1}. 
 In minisuperspace one can mimic some popular gauge choices of standard inflationary cosmology, 
\begin{align} \label{gauge1}
{\rm Unitary \ gauge:}&\qquad \qquad \qquad  \delta \phi = 0, \qquad \delta \rho =  \zeta  \qquad \qquad \\
{\rm Spatially \ flat \ gauge:}& \qquad \qquad  \qquad \delta \phi = \varphi, \qquad \delta \rho = 0\, , \label{gauge2}
\end{align}
and some of the features of the ``true" curvature perturbations $\zeta(t,\vec x)$ on super-Hubble scales are indeed  replicated at the minisuperspace level (we comment further on this while discussing the unitary gauge in Sec.~\ref{sec_pert}).
These two gauges correspond to foliating the field space at constant $\phi$ and $\rho$ respectively.

\begin{figure}[h]
\vspace{1cm}
\begin{center}
    \includegraphics[width=10cm]{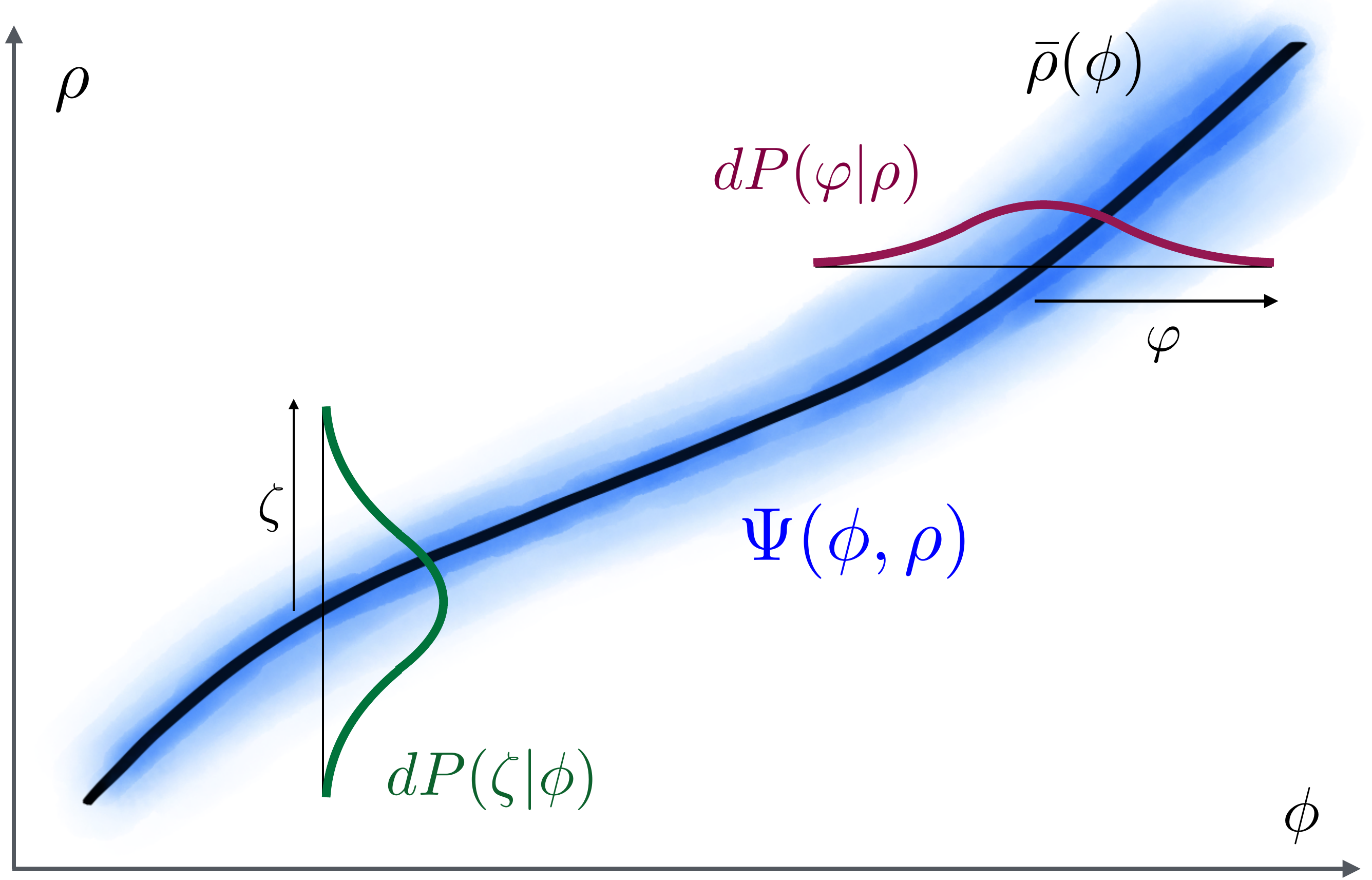} 
  \end{center}\vspace{-.4cm}
  \captionsetup{width=.95\linewidth}
\caption{\small  The WdW wavefunction $\Psi (\phi,\rho)$ (in blue) is approximately centered around the classical trajectory $\bar \rho(\phi)$. In order to obtain a probability from it one has to choose how to section the field space, \emph{i.e.} which variable should be used as ``time".  If we use $\phi$, we are going to calculate the conditional probability $dP(\zeta|\phi)$ (in green) of getting some value of $\rho$, or of $\zeta = \rho - \bar \rho(\phi)$, \emph{given that} the scalar field evaluates $\phi$. This is the unitary gauge in perturbation theory. The main alternative is using $\rho$ as time, which defines the \emph{spatially flat gauge} (in red).   } \label{fig_1}
  \end{figure}

\section{The WdW viewpoint}\label{sec_WdW_ViewPoint}
Another way of dealing with the system~\eqref{intro_squareroot} is to directly calculate the canonical momenta for each of the variables $q^\mu$, 
\begin{equation}
p_\mu = \frac{1}{\alpha H_{\star}N} g_{\mu \nu} \dot q^\nu\, . 
\end{equation}
and write the Hamiltonian before substituting the constraint, 
\begin{equation} \label{hamiltonianwdw}
 \, {\cal H} = N\, \frac{H_{\star}}{\alpha} \left( \frac{\alpha^2}{2} \, p_\mu \, p^\mu  + U (q^\mu)\right)\, . 
\end{equation}
Variation with respect to $N$ gives, simply, $ {\cal H}=0$, or its quantum version, the WdW equation
\begin{equation} \label{wdw}
\setlength{\fboxsep}{2\fboxsep} \boxed{\left(- \frac{\alpha^2}{2} \mybox + \,  U (q^\mu) \right) \Psi (q^\mu) = 0\, .}
\end{equation}

In the above, $\mybox\,\, \equiv\frac{1}{\sqrt{-g}}\partial_{\mu}\left(g^{\mu\nu}\sqrt{-g}\partial_{\nu}\right)$ is the covariant d'Alambert operator in the $d+1$ dimensional field space.
This is the only choice for $ p_\mu p^\mu$ which seems to make sense---any other choice would not be invariant under field redefinitions (see also~\cite{Nicolis:2022gzh}).  

Eq.~\eqref{wdw} is clearly different from the equation governing the perturbative wavefunction~\eqref{sch_sec2} and we do not expect the exact same physics from the two approaches. However, they should share the same classical limit. 
We tend to associate a classical behavior to a quantum system when the wavefunction consists of a rapidly oscillating phase, $
\Psi \sim e^{i S}$.
This way, the value of the momentum is highly correlated with the position, 
\begin{equation}
p_\mu  \Psi = - i \partial_\mu  \Psi \sim (\partial_\mu  S) \, \Psi \, ,
\end{equation}
which makes us identify  $S$ with the Hamilton's principal function. As in classical statistics, the system is most surely on some classical path where $p_\mu  \sim \partial_\mu  S(q)$, except that we do not know which one! In other words, despite the strong correlation position-momentum, the position itself is completely undetermined at this stage. We should thus\footnote{A nice recent review of classicality criteria is found in~\cite{Khan:2023ljg}}  equip $\Psi$ with a slowly-varying modulation factor $\psi$, 
\begin{equation} \label{form}
\Psi =  \, e^{i S/\alpha} \, \psi \, .
\end{equation}
For convenience, in the above we have redefined $S$ by including a factor of $1/\alpha$. When we stick~\eqref{form} into~\eqref{wdw}
we obtain terms containing different powers of  $\alpha$.  
At zeroth order in $\alpha$ we get the Hamilton Jacobi (HJ) equation for $S$, 
\begin{equation} \label{HJ}
\setlength{\fboxsep}{2\fboxsep} \boxed{\partial_\mu S\partial^\mu S  + 2 U =0\, .}
\end{equation}

Once $S$ is found, all remaining terms in the WdW equation compose a Schr\"odinger-like equation for $\psi$,
\begin{equation} \label{psipsi}
\setlength{\fboxsep}{2\fboxsep} \boxed{  i \left(\partial^\mu S  \partial_\mu \psi + \frac12 \mybox \! S \cdot \psi  \right) = - \frac{\alpha}{2}\, \mybox \!  \psi    \, .}
\end{equation}
Obviously, the above procedure is just the WKB approximation in more than one dimension~\cite{VanHorn:1967zz} applied to eq.~\eqref{wdw}.

\subsection{Going with the (classical) flow}\label{classflow}

A very elegant way to set boundary conditions for this system of equations is the no-boundary proposal~\cite{Hawking:1981gb,Hartle:1983ai}\footnote{See also~\cite{Janssen:2020pii,Lehners:2023yrj} for a useful summary and a comprehensive bibliography of this topic} . 
To leading order in $\alpha$ this corresponds to finding a classical complex saddle that describes a smooth euclidean geometry at early times. The corresponding $S$ contains  an imaginary part that goes to zero in the large field limit where the metric  becomes Lorentzian (e.g.~\cite{Maldacena:2024uhs}). 
 Here we are not concerned with the initial conditions of the universe. We just aim to reproduce, well within the Lorentzian regime, the closest quantum analog $\Psi$ to some classical background solution $ {\bar q}^{\, \mu}(t)$. For this reason, we take $S$ to be real.

The solutions of~\eqref{HJ} are far from unique (in the case of the Klein Gordon equation this is shown explicitly in App.~\ref{sec_HJ}).
One concrete way to compute $S$ that fits our purposes is the following.\footnote{The value of the Hamilton principal function $S(q^\mu)$ is known to  coincide with the action functional $I[q^\mu(t)]$ evaluated on a solution of the e.o.m. This can be used to produce solutions of the HJ equation (e.g.\cite{Janssen:2020pii,Lehners:2023yrj}). However, it is not clear to us if and how this trick can be used to generate \emph{any} solution of the HJ equation.} One chooses a (surface-orthogonal) congruence of classical solutions $q^{\, \mu}_{\rm cl}(t)$  of which $ {\bar q}^{\, \mu}(t)$ is an element (see Fig.~\ref{fig-1} in App~\ref{sec_HJ} for examples). By comparing~\eqref{HJ} with the classical constraint~\eqref{constraineq}, one sees that inside a ``tube"  around $ {\bar q}^{\, \mu}(t)$ one can  set 
 \begin{equation} \label{matching}
 \partial^\mu S = \dot { q}^{ \, \mu}_{\rm cl} /H_\star\, .
 \end{equation}
Not surprisingly, one can build a Fermi coordinate system around $ {\bar q}^{\, \mu}(t)$, and $S$ satisfies, locally, along the curve, an equation similar to Raychaudhuri's (see App.~\ref{fermi}). 

On the LHS of~\eqref{psipsi}, $\psi$ is hit by the vector field $\partial^\mu S \partial_\mu$. It is useful to see this term in a coordinate system adapted to the classical solution $ {\bar q}^{\, \mu}(t)$. By keeping $q^0$ as ``time", we can introduce, as in~\eqref{varphis}, spatial coordinates $\varphi^i$ that vanish on the classical solution $\bar q^{\, \mu}(t)$. However, there is a more special set of coordinates, $\zeta^i$, that Vilenkin calls \emph{comoving} in~\cite{Vilenkin:1988yd}, with the additional property to be classically conserved. In other words,  $\dot \zeta^i = 0$ is always a solution of the e.o.m., which, in the Lagrangian for the perturbations, is tantamount to the absence of any potential term. Curves of constant $\zeta^i$  represent alternative trajectories close to $\bar q^{\, \mu}(t)$ (c.f. Fig.~\ref{fig-1} in App~\ref{sec_HJ}). The curvature perturbation on comoving surfaces $\zeta_k$ in the limit $k\rightarrow0$ has this property, which motivates this choice of notation.

As $\dot \zeta^i_{\rm cl} = 0$ by construction, it is clear from~\eqref{matching} that in these coordinates the only non vanishing component of $\partial^\mu S$ is $\partial^0 S$. By using the classical solution $ { q}_{\rm cl}^{\, 0}(t)$ to convert  $q^0$ into the cosmic time $t$ we obtain
\begin{equation}\label{matching2}
\partial^\mu S \partial_\mu \psi  \, =\, H_\star^{-1} \partial_t \psi \, .\quad \ \  \quad \qquad \qquad \qquad \qquad {\rm (comoving \ coordinates}\ \zeta^i  {\rm)}
\end{equation}

Other choices of coordinates $\varphi^i$ do not enjoy classical conservation and thus appear in the quadratic Lagrangian with a potential term (this is the general situation,~\eqref{sch_sec2}). The spatial components of $\partial^\mu S$ in these coordinates vanish only on the trajectory. In this case, the vector field $\partial^\mu S \partial_\mu$ contains spacial derivative terms (at least-) \emph{linear} in $\varphi^i$. 
Schematically, 
\begin{equation} \label{genericco}
\quad \ \ \ \partial^\mu S \partial_\mu \psi  \, \simeq\, H_\star^{-1} \partial_t \psi + \# \, \varphi^i \partial_{\varphi^i} \psi \, .\qquad \qquad \qquad {\rm (generic \ coordinates}\ \varphi^i  {\rm)}
\end{equation}

\subsection{Degrees of freedom, probabilities etc.} \label{probabilities}

In a standard quantum mechanical system there are as many degrees of freedom as the number of independent variables inside the wavefunction \emph{at a given time} (the Schr\"odinger picture is assumed throughout). 
The WdW equation~\eqref{wdw} constrains the shape of $\Psi$ \emph{in the absence} of time. In order to match the standard counting, one should, roughly speaking, subtract one dimension from the total configuration space. The idea is that one of the fields, say  $q^0$, should be used as a time variable. After this choice is made,  we are free to decide  ``initial conditions" for the remaining fields on some $q^0 = { const.}$ surface in field-space.\footnote{WdW features second order derivatives in all variables, so the initial conditions on the $q^0 = { const.}$ surface would, strictly speaking, include the $q^0$-derivative of $\Psi$ on that surface. As sketched in Sec.~\ref{highertime}, however, the second $q^0$-derivatives can be seen as terms of higher order in $\alpha$ in the semiclassical expansion, so they can be dropped in a first approximation. }

These considerations are inevitably related with the probability interpretation of the wavefunction. As mentioned, the natural interpretation of the standard wavefunction $\psi_P(\varphi^i | t)$ of the perturbations is that of a \emph{conditional probability amplitude}---of finding the system at some coordinate-point $\varphi^i$ \emph{given that} the time variable evaluates $t$. 
Accordingly, the normalization of $\psi_P$ does not involve integrating over $t$ and it is in fact preserved through the $t-$evolution. 

In the case of WdW one is forced to a relational interpretation where one combination of the variables $q^\mu$ is used as time and the remaining ones are the true dynamical degrees of freedom. Accidentally, WdW, as Klein Gordon, enjoys a conservation law \emph{internal} to the configuration space, 
\begin{equation}
j^\mu \equiv \frac{i}{2}\sqrt{-g}\left(\Psi \partial^\mu \Psi^* - \Psi^* \partial^\mu \Psi\right), \qquad \partial_\mu j^\mu = 0\, .
\end{equation}
The above expression can be used to define a \emph{conditional probability}. By taking $q^0$ as the internal time for simplicity,  we should thus interpret $j^0$ as a probability density, which is conserved during time evolution. 

This interpretation is famously plagued by the fact that $j^0$ is not positive--definite. Most likely, this issue should be addressed in the context of a third quantization (\emph{i.e.} like for the solutions of the Klein Gordon equation, one should see $\Psi$ as a field, not as a wavefunction). At any rate, close to the semiclassical regime we are interested in, $j^0$ \emph{is} positive. By expressing it in terms of $S$ and $\psi$ we obtain 
\begin{equation} \label{probabilityconserved}
dP(q^i|q^0) = \sqrt{-g}\left[\frac{1}{\alpha} \partial^0 S  \left|\psi \right|^2 + \frac{i}{2}\left(\psi \partial^0 \psi^* - \psi^* \partial^0 \psi\right)\right] d \vec q\, .
\end{equation}
To leading order in $\alpha$ this quantity is always positive (or always negative, in which case we should change the sign on the RHS) as long as  $\partial^0 S$ does not switch sign along the trajectory, which would constitute a breakdown of the semiclassical approximation (see also~\cite{Vilenkin:1988yd}, and~\cite{Chataignier:2020fys,Chataignier:2020fap} on the conditional probability interpretation).

\subsection{Apples to apples}\label{apples}

A direct comparison between the probability density in perturbation theory~\eqref{probapp_intro} and that of WdW~\eqref{probabilityconserved} in the semiclassical limit $\alpha \rightarrow 0$ suggests the following relation between the two wavefunctions,\footnote{See also~\cite{Chataignier:2022iwb} for a version of this relation that does not include  the phase factor $e^{i f/\alpha}$.}
\begin{equation} \label{relationPW}
\setlength{\fboxsep}{3\fboxsep} \boxed{ \psi_P \, = \ e^{i f(q)/\alpha} \left(\sqrt{-g} \, \partial^0  S\right)^{1/2} \psi  +{\cal O}(\alpha)\, .}
\end{equation}
In Sec.~\ref{slowroll} we verify the accuracy of this relation in the case of a slow-roll scenario. 

Since both $\psi_P$ and $\psi$ appear only through their squared moduli in the expression of the probability, it is not surprising that they are related by a field dependent phase $f(q)$ in~\eqref{relationPW}. Attached to this phase there is a solution of a little puzzle, which is the following. 
 
As mentioned, in a general coordinate system $\varphi^i$  the quadratic Lagrangian, and thus the Schr\"odinger equation of perturbation theory, feature a potential term.  However, eq.~\eqref{psipsi} does not seem to be able to accommodate any potential. In fact, in the semiclassical expansion of the WdW equation, the potential has been already taken care of entirely by the HJ equation~\eqref{HJ}.  How can we then hope to match  the results of perturbation theory?

\subsubsection*{Phases and canonical transformations: a free particle example}

Let us look at the problem in the simplest possible terms. 
Let's consider the Lagrangian of a single-free particle in standard quantum mechanics, 
\begin{equation}
    L^{(1)}=\frac{1}{2} \dot{\zeta}^2\label{eq:ACTIONfreeparticle}\, .
\end{equation}
We are already in comoving variable as $\dot \zeta=0$ is always a solution of the classical equations of motion. 
The Schrödinger equation reads
\begin{equation}\label{sch1_equation_freepartice}
    i\partial_t\psi_P = - \frac{1}{2}\partial_\zeta^2\psi_P.
\end{equation}
 A less fortunate choice of position variable is  
 $\varphi \equiv \zeta \, g^{-1}(t) $, with $g$ some function of the time, in terms of which the Lagrangian becomes
\begin{equation} \label{L1}
      L^{(1)} =\frac{1}{2}\Big(g^2\dot{\varphi}^2+2\dot{g}g\, \varphi\dot{\varphi}+\dot{g}^2 \varphi^2\Big).\\
    \end{equation}
After integrating by parts we obtain
\begin{equation} \label{L2}
 L^{(2)}  = \frac{1}{2} \, \Big( g^2 \dot{\varphi}^2- g \ddot{g}\, \varphi^2\Big).
\end{equation}
The seemingly innocent integration by parts has defined a new Lagrangian $L^{(1)}\rightarrow L^{(2)}$ and thus a new conjugate momentum, related to the original one by means of a \emph{canonical transformation}. At the level of the wavefunction written in position representation this corresponds to a multiplication by a phase (see e.g.~\cite{Bozza:2003pr}). From the point of view of the path integral this is clear. The Lagrangians $L^{(1)}$ and $L^{(2)}$ differ by a total derivative which, integrated inside the action, simply multiplies the final state by a phase. More explicitly, the Hamiltonians produced by~\eqref{L1} and~\eqref{L2} give two different Schr\"odinger equations,
\begin{align} \label{psi1}
i\left(\partial_t - \frac{\dot g}{g}\, \varphi \, \partial_\varphi\right) \psi_P^{(1)} &= - \frac{1}{2 g^2} \partial_\varphi^2 \psi_P^{(1)}\, , \\[2mm] \label{psi2}
 i\, \partial_t \psi_P^{(2)} &= - \frac{1}{2 g^2} \partial_\varphi^2 \psi_P^{(2)} + \frac{g \ddot{g}}{2}\, \varphi^2\psi_P^{(2)}\, .
\end{align}  
One can check that $\psi_P^{(1)}$ and $\psi_P^{(2)}$
 are related by
\begin{equation}
\psi_P^{(2)}(\varphi,t) = g^{\frac{1}{2}}\exp\left(- \, i\, \frac{ g \dot g}{2}\, \varphi^2 \right)\, \psi_P^{(1)}(\varphi,t).
\end{equation}

Notice that~\eqref{sch1_equation_freepartice} and~\eqref{psi1} is the same differential equation (\emph{i.e.} the same differential operator equated to zero), just written in different (configuration space + time) coordinates $(t, \zeta) \rightarrow (t,\varphi )$. On the other hand, between~\eqref{psi1} and~\eqref{psi2} there is a (canonical) transformation of momenta and Hamiltonians, which implies a redefinition of the wavefunction. The spatial ($\varphi$-) derivative term on the LHS of~\eqref{psi1} is, upon a canonical transformation, completely equivalent to a \emph{potential} term, i.e. the second term on the RHS of~\eqref{psi2}.  

This is of course relevant to our setting because the equation for $\psi$,~\eqref{psipsi}, cannot contain a potential term. 
However, once expressed in some generic perturbation variables $\varphi^i$, terms of the type ``$ i \varphi \partial_\varphi \psi$" are included in this equation, as discussed at length around Eq.~\eqref{genericco}. In order to reabsorb these terms we need to multiply the wavefunction by a phase. This procedure fixes the phase $f$ in~\eqref{relationPW} univocally and, at the same time, it recovers the correct potential term of perturbation theory. This can be all seen explicitly in the spatially flat example of Sec.~\ref{slowroll}.

\subsection{Higher time derivatives} \label{highertime}

We have seen that perturbation theory  and  WdW equation match in the semiclassical limit ($\alpha \rightarrow 0$), if the two wavefunctions are related as in~\eqref{relationPW}, and if we use the rules for the probabilities~\eqref{probapp_intro}  and~\eqref{probabilityconserved}. It also follows quite trivially that no deviation from the classical behavior is expected in the semiclassical limit. This is best seen by using the Schrodinger equation of perturbation theory, that 
in comoving coordinates $\zeta^i$ and for $\alpha=0$ simply reduces to $i \partial_t \psi_P = 0$. One can initially prepare the system in some smooth normalizable state $\psi_P(\zeta^i)$. The wavefunction and its associated probability is simply constant in time, so if initially peaked at $\zeta^i = 0$, it remains peaked there at any time. 

The diffusive terms of ${\cal O}(\alpha)$ lead to deviations from the classical behavior, as well as to different predictions between the perturbative and the WdW approaches. The distinctive feature of WdW is the presence of second order derivatives with respect to all dynamical fields present, including the field that is going to be used as time. In fact, eq.~\eqref{psipsi} contains a d'Alambertian operator $\mybox$ on the RHS, while the equation for $\psi_P$,~\eqref{sch_sec2}, features a spatial Laplacian. 
When one obtains the Schr\"odinger equation as the non-relativistic limit of the Klein Gordon (KG) equation, one finds that  higher $t$-derivatives are suppressed by higher powers of the speed of light $c$. Similarly, one can see here that higher time derivatives are effectively suppressed by higher powers of $\alpha$. 

We show this heuristically in a  simplified 1+1 dimensional setup. We assume a Gaussian ansatz for $\psi$, centered around the classical trajectory. In the comoving coordinate $\zeta$, 
\begin{equation} \label{gaussianansatz}
\psi(\zeta , q^0) \ = \ {\cal N}(q^0) \exp\left[- \frac12 \Omega(q^0) \zeta^2 \right]\, ,
\end{equation}
where some field $q^0$ has been chosen as time ($\partial_0 \equiv \partial_{q^0}$ in what follows).
By sticking this ansatz inside a simplified version of~\eqref{psipsi}, 
\begin{equation}
i \partial_0 \psi = - \frac{\alpha}{2}\left( \partial_\zeta^2 + B \, \partial_0^2\right)\psi\, ,
\end{equation}
for some constant $B$, one can regroup terms of different orders in $\zeta$. To zeroth order one gets an equation for the normalization factor, 
\begin{equation} \label{enne}
\partial_0 {\cal N} = {\cal O}(\alpha)\, .
\end{equation}
The terms proportional to $\zeta^2$, equated to zero, give 
\begin{equation} \label{ric}
i \partial_0 \Omega = \alpha \Omega^2 - \alpha \frac{B}{2}  \partial_0^2 \Omega + {\cal O}(\alpha^2)\, ,
\end{equation} 
where~\eqref{enne} has been used. 

We can imagine a smooth situation where the variance of $\zeta$ gently evolves along $q^0$ like in Fig.~\ref{fig_1}. In this case we could attempt to solve for $\Omega$ around a constant solution, plus $\alpha$-corrections that slowly evolve in time. By inserting this type of ansatz inside~\eqref{ric} one sees that the term proportional to $B$ gives a contribution only to order $\alpha^2$.

\subsection{Non-gaussian terms and deviations from the classical behavior} \label{deviat}

The WdW approach, as we are seeing, is certainly more convoluted than perturbation theory. However, it naturally conveys an 
 ``exact" result, in the sense that all non-linear terms are automatically included in the WdW equation for $\psi$~\eqref{psipsi}. 
 This formalism could thus be used to investigate effects (e.g. primordial black holes formation) that are hidden in the tail of the probability distribution~\cite{Celoria:2021vjw,Creminelli:2024cge}. 
 
Another possible matter on which the WdW viewpoint could be of help is the study of the deviations from the classical background solutions. 
Quantum  averages evolve in time differently than classical solutions, in general. The Ehrenfest theorem, for a particle at position $x$ in a potential $V(x)$, famously implies 
\begin{equation} 
 \langle \ddot x \rangle  = -\,  \langle V'( x)\rangle\, ,
\end{equation}
Crucially, $\langle V'( x)\rangle \neq V'(\langle x\rangle)$ in general. Depending on its spread, the wave-function can ``feel" regions of the potential away from its peak. Quantum tunneling is the most spectacular display of this phenomenon, but milder deviations from the classical behavior are ubiquitous.

One would expect to see these deviations also in perturbation theory and this is certainly the case for ordinary (finite-dimensional) quantum mechanical systems. If we expand around a classical solution $\bar x(t)$ by defining $\delta x = x - \bar x(t)$, the action for the perturbations  will generally take the form
\begin{equation}
I = \frac12\int dt \ A(t) \dot {\delta x}^2 - B(t) {\delta x}^2 + C(t) {\delta x}^3 + \dots\, .
\end{equation}
As one tries to evolve e.g. an initially Gaussian wavepacket with the Hamiltonian derived from this action one finds that the interaction term proportional to $C(t)$ not only generates non-gaussianity, it also shifts the peak of the wavepacket away from $\delta x =0$. In other words, $\langle \delta x(t) \rangle = 0$ is inconsistent with the dynamics. This is how perturbation theory ``sees" that $\langle x(t) \rangle \neq \bar x (t)$. 

In standard \emph{cosmological} perturbation theory, however, averages of perturbations are always zero by construction. The expectation value of a Fourier mode, $\langle \zeta_k\rangle$ vanishes simply because of translational invariance. However, deviations from the classical solutions are clearly seen at the mini-superspace level, both in the WdW framework and in perturbation theory where they are conceived, as we have seen, as non-gaussianity. What sense should be made of these deviations? Again, mini-superspace results apply most straightforwardly to the zero mode of a spatially closed universe.  
 In this context, deviations from classicality have already been noticed in~\cite{Lehners:2024qaw}. 
However, by reasoning along the  lines  of  Sec.~\ref{minisuper},  one could argue that these deviations could also be a generic feature of any super-Hubble patch as it causally detaches from the rest of the universe. 

How large is the effect? It is suppressed by the small parameter $\alpha$ defined in~\eqref{alphaa} so one has to go beyond the semiclassical limit to see it. However,  it is   enhanced during inflation by a factor of $1/\epsilon$ (see Eq. ~\eqref{wdwunitary}). So, the regime where deviations become sizable is the one of eternal inflation, which is in fact characterized by $\alpha/\epsilon \gtrsim1$. The picture (e.g.~\cite{Linde:1986fd,Linde:1986fc,Guth:2007ng}) is that during eternal inflation the fluctuations of the field compete with the classical evolution over a Hubble time, and the field effectively behaves stochastically.  Our deviations, on the other hand, are deterministic and computable, given some initial condition.  So they seem to \emph{add} to the standard stochastic picture rather than rephrase it.

\section{Perturbation theory for a scalar field} \label{dom_component}

In order to apply the formalism that we have discussed we work in mini-superspace with a metric of the form
 \begin{equation}
 ds^2 = - N^2 dt^2 + e^{2\rho(t)} d\vec x^{\, 2}\, ,
 \end{equation}
and model the matter content with a scalar field $\phi(t)$.  The action reads 
\begin{equation}
    I=\frac{1}{\alpha H_{\star}} \int dt \, e^{3\rho}\bigg(-\frac{3\dot{\rho}^2}{N}+\frac{\dot{\phi}^2}{2N}-NH_\star^2V(\phi)\bigg).\label{eq:action_model1} 
\end{equation}
Notice that $\phi$ is dimensionless (\emph{i.e.} $\phi = \hat \phi/M_P$ with $\hat \phi$ a canonical scalar field). 
We have in mind a classical solution such that $H_\star^2\simeq\dot{\rho}^2(t_\star)$ at a reference time $t_\star$, at which the field satisfies $\phi(t_\star)=0$. The minisuperspace metric associated to~\eqref{eq:action_model1} is $g_{\mu\nu}=e^{3\rho} \diag(-6,1)$.
%\begin{equation}  g_{\mu\nu}=e^{3\rho}\left(\begin{array}{cc}
%-6 & 0 \\
%0 & 1
%\end{array}\right)\, .
%\end{equation}

Varying with respect to $N$, $\rho$ and $\phi$ and then setting $N=1$ gives the classical Hamiltonian constraint and the two equations of motion,
    \begin{subequations}\label{eqs:class_sol_scalar_field}
    \begin{align}
        3 \dot{\rho}^2&=\frac{\dot{\phi}^2}{2}+H_\star^2V(\phi),\label{eq:hamiltonian_constraint_model1}\\
        \ddot{\rho}&=-\frac{1}{2}\dot{\phi}^2,\label{snd_class_eom}\\
\ddot{\phi}&+3H\dot{\phi}+H_\star^2V'(\phi) = 0 \end{align} \end{subequations}

These equations are shift-invariant under $\rho\rightarrow \rho + \rho_0$. This can be used to produce a congruence of classical solutions once we have found one. 
Despite involving two fields, the scalar-field coupled to the FLRW-gravity system possesses only one dynamical degree of freedom.  This can be understood by perturbing around a classical background solution of eqs.~\eqref{eqs:class_sol_scalar_field}.

\subsection{Perturbations} \label{sec_pert}

Distilling the unique scalar degree of freedom of single field inflation is by now a standard calculation (e.g.~\cite{Garriga:1999vw,Maldacena:2002vr}). Here we are dealing with a particularly simple system 
because we are working, strictly, at zero momentum. 

Variation of Eq. \eqref{eq:action_model1} w.r.t. $N$ gives the constraint, 
\begin{equation} \label{N}
N = \sqrt{\frac{3 \dot{\rho}^2- \frac{1}{2}\dot \phi^2 }{ H_\star^2 V(\phi)}}\, .
\end{equation} 
By substituting this back into the action we obtain
\begin{equation}
    I=-\frac{2}{\alpha}\int dt e^{3\rho}\sqrt{V(\phi)\left(3\dot{\rho}^2 - \frac{1}{2}\dot{\phi}^2\right)}\, .\label{eq:action_model1_OSN}
\end{equation}

One is free to choose the value of $N$, as this corresponds to choosing a time coordinate. Small perturbations around a solution $\bar \rho(t)$, $\bar \phi(t)$  transform, under an infinitesimal time-diffeomorphism $t \rightarrow t+ \delta t \,$, as 
\begin{align} 
\delta \rho & \rightarrow \delta \rho + \dot {\bar \rho} \, \delta t \, , \\ 
\delta \phi & \rightarrow \delta \phi + \dot {\bar \phi} \, \delta t \, .
\end{align}
Clearly, by appropriately rescaling time, one can set either \(\delta \phi = 0\) or \(\delta \rho = 0\), which defines the \textit{gauge choice}. The classical background solutions are denoted by \(\bar{\rho}\) and \(\bar{\phi}\). However, to simplify the notation, in this section we will drop the overbar symbol and \(\rho\) and \(\phi\) should be understood as background quantities.

\subsubsection*{Unitary gauge}

It is customary to call unitary gauge that gauge choice in which most degrees of freedom are encoded in the gauge fields, in this case in the metric. This corresponds to setting
\begin{equation} \label{def_zeta}
   \delta\phi = 0, \quad \delta \rho=  \zeta,
\end{equation}
which defines the variable $\zeta$. 
By expanding directly inside Eqs. \eqref{N} and \eqref{eq:action_model1_OSN} one sees that this gauge choice can be enforced by chosing  \(N = 1 + \delta N\), 
with 
\begin{equation}
    \delta N = \frac{3\dot{\rho}}{H_\star^2 V(\phi)}\dot{\zeta} ,
\end{equation}
and the action for $\zeta$ reads, up to quadratic order, 
\begin{equation} \label{I1zeta}
    I_{\zeta} = \frac{3}{2\alpha H_{\star}} \int dt \, \frac{e^{3\rho}\dot\phi^2}{H_\star^2V(\phi)} \, \dot{\zeta}^2.
\end{equation}
The Hamiltonian for this model is simply  
\begin{equation}
 H_{\zeta}=\alpha H_{\star}\frac{e^{-3\rho}}{6}\frac{H_\star^2V(\phi)}{\dot{\phi}^2}p_\zeta^2\,.
 \end{equation}
 By expressing the conjugate momentum to $\zeta$ as $p_\zeta = - i \partial_\zeta$, one gets to the following Schr\"odinger equation
\begin{equation} \label{eq_sch}
    i  H_{\star}^{-1}\partial_t \psi_P = - \alpha\frac{e^{-3\rho}}{6}\frac{H_\star^2V(\phi)}{\dot{\phi}^2} \partial_\zeta^2  \psi_P\,.
\end{equation}

The behavior of this minisuperspace variable $\zeta(t)$ bears similarities with the famous $\zeta(\vec x,t)$ variable of cosmological perturbation theory, which in Fourier space (and for speed of sound $c_s =1$) is governed by the action~\cite{Garriga:1999vw,Maldacena:2002vr}
\begin{equation}
I[\zeta_k] = \frac12\int dt \, d^3k \, \frac{e^{3 \rho} {\dot \phi}^2}{{\dot \rho}^2}\left(\dot \zeta_k \dot \zeta_{-k} - e^{- 2 \rho} k^2 \zeta_k \zeta_{{\rm-}k} \right)\, .
\end{equation}
By comparing this with~\eqref{I1zeta} we see that the ``pump field" in front of the respective kinetic terms is essentially the same. In fact, $\dot \rho^2\, \propto \, V$ both in slow-roll and along the scaling solutions of constant equation of state. 
In virtue of this, the classical behavior of $\zeta_k$ on super-Hubble scales ($k \ll \dot \rho$) is analogous to that of $\zeta$.\footnote{The relation between these two variables is further studied in~\cite{HNP}. We thank the authors for discussions and for sharing a preliminary version of their draft.}  Quantum mechanically their variances behave in the same way. However, the imaginary parts of their wavefunctions behave differently in that $\zeta$ does not become as squeezed as $\zeta(\vec x,t)$. This will be detailed elsewhere. 

\subsubsection*{Spatially flat gauge}
Setting instead $\delta \rho = 0$ is reminiscent of the cosmological \emph{spatially flat gauge}, where one choses the time coordinate in such a way to carve $t = const.$ spatial surfaces that are intrinsically flat. 
 The degree of freedom is thus encoded in the scalar field, 
\begin{equation} \label{defvarphi}
  \delta\phi = \varphi, \quad \delta \rho = 0\, .
\end{equation}
This gauge is enforced by the choice 
\begin{equation}
\delta N = - \frac{\dot{\phi} \dot \varphi + H_\star^2V'(\phi) \varphi}{2H_\star^2V(\phi)}\,.
\end{equation}
The action for $\varphi$ then reads 
\begin{equation} \label{I1varphi}
    I_{\varphi} =\frac{3}{2\alpha H_{\star}} \int dt  \, \frac{\dot{\rho}^2e^{3\rho}}{H_{\star}^2V(\phi)} \left[\dot{\varphi}^2 - H_\star^2 \left( V''(\phi) -\frac{V'(\phi)^2}{V(\phi)}  \right) \varphi^2   \right],
\end{equation}
from which we can derive the associated Schrödinger equation following the same procedure as in the unitary gauge. It reads
\begin{equation}\label{eq:Scheq_Pert_sfg}
    i H_{\star}^{-1}\partial_t\psi_P=-\frac{\alpha}{6}\frac{H_{\star}^2V(\phi)}{\dot{\rho}^2e^{3\rho}}\partial_{\varphi}^2\psi_P-\frac{3}{2\alpha}\frac{\dot{\rho}^2e^{3\rho}}{
    H_{\star}^2V(\phi)}\left(\frac{V'(\phi)^2}{V(\phi)}  -V''(\phi)\right)\varphi^2\psi_P\,.
\end{equation}
One clearly sees that, as opposed to unitary gauge, the spatially flat gauge generally features a mass term. 

\subsection{Scalar field models}

We concentrate our study on two specific potentials \( V(\phi) \) of interest, defining each a different model.

\subsubsection*{Model 1: Exponential potential}The first is an exponential potential 
\begin{equation}\label{eq:potential_scalingsol}
V(\phi) = (3 - \epsilon)e^{-\sqrt{2\epsilon}\phi},
\end{equation}
which allows attractor scaling solutions of constant slow-roll parameter $\epsilon$, 
    \begin{equation}
     \rho(t)=\frac{1}{\epsilon}\ln\frac{t}{t_\star},\qquad \phi(t)=\sqrt{\frac{2}{\epsilon}}\ln\epsilon H_\star t. \label{eq:classical_solutions_model1}
\end{equation}
Here \emph{scaling} refers to the property of this classical solution of maintaining a constant equation of state so that the kinetic and potential components remain in the same proportions during time evolution. 
With an appropriate choice of the constant $t_{\star}$ we can express the background solution above in a non-parametric way,
\begin{equation}\label{eq:Scaling_Solutionn}
 \rho(\phi) = \frac{\phi}{\sqrt{2\epsilon}}\, .
\end{equation}
The slow-roll parameter $    \epsilon\equiv- \frac{\dot H}{H^2} $
here does not need to be small  (the model could be used to mimic some dominant component also during deceleration, e.g. $\epsilon = 2$ and $\epsilon = 3/2$ during radiation and matter dominance respectively). We find it convenient to use it   instead of $w\equiv p/\rho$.  On the scaling solution we indeed have the relation between $\epsilon$ and $w$ being given by $\epsilon = \frac{3}{2}(w+1)$. 

In Sec.~\ref{scaling} we apply the WdW equation to this potential. Because of the scaling behavior, this is a highly symmetric model. One sees that the combination of derivatives of $V$ that gives rise to a mass term for $\varphi$ in the spatially flat gauge (eq.~\eqref{I1varphi}) vanishes in this case. In order to have a non-vanishing mass in perturbation theory, and verify the general formalism of Sec.~\ref{apples}, we need to extend the pure exponential model, for example by adding a non-vanishing $\eta$ slow-roll parameter.

\subsubsection*{Model 2: Slow-roll }
The second potential allows both slow-roll parameters  
\begin{equation}
    \epsilon(\phi) \equiv -\frac{\dot{H}}{H^2},\quad \eta(\phi) \equiv \frac{\dot{\epsilon}}{\epsilon H},
\end{equation}
to be nonzero. In the regime where they are both small, one finds
\begin{equation}
    \epsilon(\phi) = \epsilon + \sqrt{\frac{\epsilon}{2}}\eta\, \phi + \dots\, ,
\end{equation}
where, here and throughout, $\epsilon \equiv \epsilon(0)$ and $\eta \equiv \eta(0)$. Using the classical Eqs.~\eqref{eqs:class_sol_scalar_field} one also finds the relation between the kinetic term and the potential
\begin{equation}
    V (\phi)= \frac{3 - \epsilon(\phi)}{\epsilon(\phi)}\frac{\dot{\phi}^2}{2}.
\end{equation}
This allows to build the potential with the following expansion at the exponent
\begin{equation} \label{pottential_SR}
V(\phi) = (3 - \epsilon) \exp\left[-\sqrt{2\epsilon} \left(1 + \frac{\eta}{6}\right)\phi - \frac{\eta}{4}\phi^2 + \dots \right].
\end{equation}
The approximate classical solutions for this model will be worked out in Sec.~\ref{slowroll}. Notice that for this model the mass term in the spatially flat gauge is proportional to \(\eta\), and the Schrödinger equation in this gauge reads 
\begin{equation}\label{eq:sch_sfg_perturbation1}
    iH_{\star}^{-1}\partial_t\psi_P=-\frac{\alpha}{6}\left[3-\left(\epsilon+\sqrt{\frac{\epsilon}{2}}\eta\rho\right)\right]e^{-3\rho}\partial_{\varphi}^2\psi_P-\frac{3e^{3\rho}\epsilon\eta}{2\alpha}\varphi^2\psi_P\, .
\end{equation}

\section{Application to Model 1} \label{scaling}

In this section we want to obtain the WdW equation to our scalar-field coupled to FLRW-gravity model in the case where the potential is given by Eq. \eqref{eq:potential_scalingsol}. No slow-roll approximation is used in this section and the limits on the parameter $\epsilon$ are just $0<\epsilon<3$.  
The HJ equation \eqref{HJ}  reads
\begin{equation}
    -\frac{1}{6}(\partial_{\rho}S)^2+(\partial_{\phi}S)^2+2(3-\epsilon)e^{6\rho-\sqrt{2\epsilon}\phi}=0,
\end{equation}
whose solution can be chosen to be
\begin{equation}
    S(\rho,\phi)=-2e^{3\rho-\sqrt{\frac{\epsilon}{2}}\phi}.
\end{equation} 
We can associate the above with the congruence of classical solutions 
\begin{equation}
 \rho_{\rm cl}(\phi) = \frac{\phi}{\sqrt{2\epsilon}} + \rho_0\, ,
\end{equation}
obtained from~\eqref{eq:Scaling_Solutionn} by applying shift symmetry. 
The d'Alembertian of $S$ reads
\begin{equation} \label{laplacianfirst}
\mybox \!S=\, (3-\epsilon) e^{- \sqrt{\frac{\epsilon}{2}}\, \phi}\,.
\end{equation}
By using the above expressions we finally get the following equation for $\psi$, 
\begin{equation} 
   i e^{-\sqrt{\frac{\epsilon}{2}}\phi}\Big(\partial_{\rho}\psi+\sqrt{2\epsilon}\partial_{\phi}\psi+\frac{3-\epsilon}{2}\psi\Big)=\frac{\alpha}{2}e^{-3\rho}\Big(\frac{1}{6}\partial_{\rho}^2\psi-\partial_{\phi}^2\psi\Big),
    \label{eq:adaptedHJ2ndmodel}
\end{equation}
which is exact, in the sense that it contains all the remaining pieces of the WdW equation after the HJ equation has been applied. We still have to choose which field, $\rho$ or $\phi$, will play the role of ``time". We expect that the first two terms on the LHS will arrange to give a time derivative, once we expand around the classical trajectory (see discussion around eqs.~\eqref{matching} and~\eqref{matching2}). The last, third term on the LHS comes instead from the d'Alembertian~\eqref{laplacianfirst}, which is only dependent on $\phi$  because of the shift symmetry in $\rho$. If we then consider $\rho$ as the dynamical variable and $\phi$ as time this will be a time dependent---but constant in space---potential term in the Schroedinger equation that will only affect the normalization of the wavefunction. This suggests the use of the \emph{unitary gauge} in which $\phi$ is treated as time. 
\subsection{Unitary gauge}
It is convenient to use the coordinates in the field space adapted to the scaling solution in Eq. \eqref{eq:Scaling_Solutionn}
\begin{equation}
\left\{
\begin{aligned}
    \zeta(\rho, \phi)&=\rho-\bar \rho(\phi), \\
    \phi(\rho, \phi)&=\phi,
\end{aligned}
\right.
\end{equation}
with $\bar{\rho}(\phi)=\frac{\phi}{\sqrt{2\epsilon}}$.

This field-space coordinate transformation induces a chain-rule transformation on the partial derivatives that should be applied on both sides of~\eqref{eq:adaptedHJ2ndmodel}. 
Notice that $\zeta$ is the same at linear order as that defined in Eq. \eqref{def_zeta}.
With these substitutions Eq. \eqref{eq:adaptedHJ2ndmodel} simplifies to 
\begin{multline}\label{finalschroed}
i  \, e^{-\sqrt{\frac{\epsilon}{2}}\phi}\left(\sqrt{2\epsilon} \partial_\phi \psi+\frac{3-\epsilon}{2}\psi\right)=\\ -\frac{\alpha}{12}\frac{3-\epsilon}{\epsilon} e^{-\frac{3}{\sqrt{2\epsilon}}\phi}(1
\color{red}-3\zeta+\dots
\color{black})\left[\partial_\zeta^2\psi\color{red}+\frac{6\epsilon}{3-\epsilon}\left(\, \partial_\phi^2\psi-\sqrt{\frac{2}{\epsilon}}\, \partial_\zeta\partial_\phi\psi\right)\color{black}\right]\, ,
\end{multline}
where the black terms match the predictions of the perturbative approach, while the red terms are new to the WdW approach. 
The non-Gaussian terms that multiply the diffusive term on the right-hand side are likely also present in perturbation theory beyond quadratic order (although we have not explicitly verified this). However, the higher time derivatives within the square brackets on the RHS are distinctly unique to the WdW equation. As discussed in Sec.~\ref{highertime}, these terms are effectively suppressed by higher powers of \(\alpha\). 

On the left-hand side of the equation, the  terms of zero-th order in $\alpha$ are recovered exactly, as expected. To verify this, one simply needs to substitute in Eq.~\eqref{eq_sch} the scaling solution for \(\phi(t)\) from Eq.~\eqref{eq:classical_solutions_model1} and Eq.~\eqref{relationPW}, using in that case that,
\[
\left(\sqrt{-g} \partial^{\phi} S\right)^{1/2} = \left(2 \sqrt{\frac{3}{\epsilon}} \left(3 - \epsilon\right) \exp\left[3 \zeta + \frac{1}{\sqrt{2 \epsilon}} \left(3 - \epsilon\right) \phi\right]\right)^{1/2}.
\]

\subsection{Spatially flat gauge}
If instead \(\rho\) is chosen as the time variable in~\eqref{eq:adaptedHJ2ndmodel} one should expand around the inverted classical scaling solution in Eq. \eqref{eq:Scaling_Solutionn}, $\bar \phi(\rho) = \sqrt{2 \epsilon}\, \rho$. In this case it is natural to introduce the perturbative variable $\varphi$ defined in~\eqref{defvarphi}, as $\varphi = \phi - \bar \phi(\rho)$.  

The Schrödinger equation in this case takes the form
\begin{multline}
ie^{\epsilon\rho}\left(\partial_{\rho}\psi+\frac{3-\epsilon}{2}\psi\right)\\=-\frac{\alpha}{6}(3-\epsilon)e^{-3\rho}\left(1\color{red}+\sqrt{\frac{\epsilon}{2}}\varphi+\dots\color{black}\right)\left[\partial^2_{\varphi}\psi\color{red}+\frac{1}{3-\epsilon}\left(\sqrt{2\epsilon}\partial_{\varphi}\partial_{\rho}\psi-\frac{1}{2}\partial_{\rho}^2\psi\right)\color{black}\right]\,.
\end{multline}
Consequently, deviations arise exclusively from non-Gaussianities not captured by the quadratic-perturbative approach. The equation does not include any mass term, as it vanishes in this specific model under perturbation theory. This contrasts to the scenario in the slow-roll Model 2, which acquires a mass term as we will now explore in detail.

\section{Application to Model 2}\label{slowroll}

In this section, we compare the perturbation-theory and WdW approaches in slow-roll. So here we assume that $\epsilon \ll 1$ and $\eta \ll 1$. The potential is specified in Eq. \eqref{pottential_SR} and, to simplify the notation, we rescale the field $\phi$ as follows \begin{equation} \chiTwo \equiv \frac{\phi}{\sqrt{2\epsilon}}\,. \end{equation}
By using $\phi$ as time (see discussion around eq.~\eqref{intro_squareroot2}) we get an equation for the classical behaviour of $\rho(\chiTwo)$, 
\begin{equation}
\rho'' - \left(1+ \frac{V'}{V} \frac{\rho'}{2\epsilon}\right)\left(3 \rho'^{\, 2} - \epsilon\right) = 0\, ,
\end{equation}
where a prime denotes derivative with respect to $\chiTwo$. We solve the above by expanding around $\chiTwo=0$ where we fix position and velocity as $\bar \rho(0) = 0$ and $\bar \rho\, '(0) = 1$. We find
\begin{equation} \label{approxxx}
\bar \rho(\chiTwo) =   \chiTwo - \frac{(3 - \epsilon)\eta}{12} \, \chiTwo^2 - \frac{\epsilon\, \eta}{12} \, \chiTwo^3 + \dots\, .
\end{equation}
We can now exploit the shift symmetry of the system  under $\rho \rightarrow \rho + \rho_0$ and write an entire  congruence of classical solutions $\bar \rho(\chiTwo) + \rho_0$. By using standard slow-roll manoeuvring one finds an approximate expression for $\dot \chiTwo$ and then, using Eq. \eqref{approxxx}, for $\dot \rho$, along the congruence of solutions, 
 \begin{align}\label{1}
 \dot  \chiTwo &= H_\star e^{- \epsilon \chiTwo}\left(1+   \frac{(3 - \epsilon) \eta}{6} \chiTwo \right)\, ,\\[2mm] \label{2}
 \dot \rho &= H_\star e^{- \epsilon \chiTwo}\left(1 - \frac{\epsilon \eta}{4} \chiTwo^2\right)\, . 
 \end{align}
A factor of $e^{- \epsilon \chiTwo}$ has been pulled out of these expressions for convenience, but these are still expansions valid at small $\chiTwo$. 
 Similarly, one can write the potential~\eqref{pottential_SR} as
 \begin{equation}\label{eq:potentialV2SR}
 V(\chiTwo) = (3 - \epsilon) \exp{\left[- 2 \epsilon \left(1  + \frac{ \eta}{6}\right)\chiTwo - \frac{\epsilon\eta}{2} \chiTwo^2\right]}\, .
 \end{equation}
 
Eqs.~\eqref{1}-\eqref{2} represent the two components of the vector tangent to some congruence $\bar q^{\, \mu}$, so we can attempt to find $S$ by posing $H_\star \partial^\mu S = \dot {\bar q}^{\, \mu}$. One can double-check that these vectors satisfy the HJ equation, 
 \begin{equation}
 - 6 \dot \rho^2 + 2\epsilon \dot \chiTwo^2 + 2 V = 0\, ,
 \end{equation}
where a factor of $e^{3 \rho} $ has been simplified from each term. Then we should check that the corresponding covariant vector field $\bar q_{\, \mu}$ is surface orthogonal, $\partial_\chiTwo \bar q_\rho - \partial_\rho \bar q_\chiTwo =0$, so that it can be written as the gradient of $S$. This is indeed the case barring terms of ${\cal O}(\epsilon^2)$. In summary, at the relevant order in slow-roll, we find 
\begin{equation}
S = - 2 e^{3 \rho - \epsilon \chiTwo}\left(1 - \frac{\epsilon \eta}{4} \chiTwo^2\right)\, .
\end{equation}
 From which one calculates 
 \begin{equation}
\mybox \!S = \left[(3 - \epsilon)\left(1+ \frac{\eta}{6}\right) - \frac{\epsilon \eta}{2} \chiTwo - \frac{3 \epsilon \eta}{4} \chiTwo^2 \right] e^{- \epsilon \chiTwo}\, .
 \end{equation}
 We have now all the ingredients to write the equation for $\psi$~\eqref{psipsi}. It reads, 

\begin{equation} \label{eq:SCH_eq_slowroll}
i \left[e^{-\epsilon \chiTwo} \left(1 - \frac{\epsilon \eta}{4}\chiTwo^2\right) \partial_\rho \psi + e^{-\epsilon \chiTwo} \left(1 +  \frac{(3 - \epsilon) \eta}{6} \chiTwo \right) \partial_\chiTwo \psi + \frac\psi2 \mybox \!S  \right] = \frac{\alpha}{2} e^{- 3 \rho} \left(\frac16 \partial_\rho^2 \psi - \frac{1}{2\epsilon}\partial^2_\chiTwo \psi\right)\, .
\end{equation}
 The above does not contain any approximation other than slow-roll and the expansion at small $\chiTwo$. It is thus valid  in the vicinity of the axis $\chiTwo = 0$. 
 \subsection{Unitary gauge}
In this gauge we use the field $\chiTwo$ as time and $\zeta =  \rho - \bar \rho(\chiTwo)$ as spatial coordinate. We thus apply on both sides of~\eqref{eq:SCH_eq_slowroll} the change of variables $(\chiTwo,\rho)\rightarrow (\chiTwo,\zeta)$. By applying the chain rules
  \begin{align}
 \partial_\rho &\rightarrow \partial_\zeta \\
 \partial_\chiTwo &  \rightarrow \partial_\chiTwo- \bar \rho'(\chiTwo) \partial_\zeta
 \end{align}
  one sees that the partial derivatives with respect to $\zeta$ on the LHS cancel exactly, 
  \begin{multline} 
i \left[  e^{-\epsilon \chiTwo} \left(1 +  \frac{(3 - \epsilon) \eta}{6} \chiTwo \right) \partial_\chiTwo  + \frac12 \mybox \!S  \right]\psi  \\[2mm] = - \alpha \ 
e^{- 3 \bar \rho(\chiTwo)} {\color{red} e^{- 3 \zeta}}\,  \left[ \frac{3-\epsilon}{12 \epsilon} \,  \partial_\zeta^2  \, {\color{red} + \, \frac{1}{4 \epsilon} \left(({\bar \rho\, '}^2 - 1)  \, \partial_\zeta^2  - \bar \rho\, '' \partial_\zeta   - 2 \bar \rho' \partial_\chiTwo \partial_\zeta + \partial_\chiTwo^2 \right)}\right] \psi\, . \label{wdwunitary}
\end{multline}
This equation descends from the WdW equation and should be compared with what is obtained in perturbation theory. From eq.~\eqref{eq_sch} by making the semiclassical ansatz 
\begin{equation}
\psi_P = \left(\sqrt{-g} \, \partial^\chiTwo  S\right)^{1/2} \psi \simeq \exp \left[\frac32(\bar \rho(\chiTwo)  - \epsilon \chiTwo) + \frac{3 - \epsilon}{12} \eta \, \chiTwo + \frac{3 \zeta}{2} \right] \psi   
\end{equation}
after some algebra we find 
  \begin{multline} 
i \left[  e^{-\epsilon \chiTwo} \left(1 +  \frac{(3 - \epsilon) \eta}{6} \chiTwo \right) \partial_\chiTwo  + \frac12 \mybox \!S  \right]\psi  \\[2mm] = - \alpha \ 
e^{- 3 \bar \rho(\chiTwo)} \,  \left[ \frac{3-\epsilon}{12 \epsilon} \, \partial_\zeta^2 \, { \color{red} +\,  \frac{3-\epsilon}{12 \epsilon}  \left(- \frac{\epsilon\eta}{3} \chiTwo \, \partial_\zeta^2 +  \left(1-  \frac{\epsilon\eta}{3} \, \chiTwo \right)  \left(  3 \partial_\zeta  + \frac94\right) \right)}\right]  \psi\, . \label{schounitary}
\end{multline}
We highlighted in red the mismatch between WdW~\eqref{wdwunitary} and the Schr\"odinger equation of perturbation theory~\eqref{schounitary}. The two equations coincide in the semiclassical limit $\alpha =0$ as they should. At order $\alpha$, however, there is no reason to expect a perfect agreement between~\eqref{wdwunitary} and~\eqref{schounitary}. Among the terms on the RHS, the main diffusive term $\partial_\zeta^2 \psi$ starts deviating at order $\eta$, in that it is multiplied in the two equations by two slightly different  ``time" ($\chiTwo$-) dependent functions. Other than this there are two main sources of disagreement 
\begin{enumerate}
\item \emph{Non-gaussian terms.} In perturbation theory we have developed the action only to quadratic order. This is most probably the reason why eq.~\eqref{schounitary} does not contain the factor of $e^{- 3 \zeta} = 1 - 3 \zeta + \dots$ on the RHS.  These terms produce non-gaussianity in the wave function as well as deviations from the classical trajectory. Such deviations are $\alpha$-suppressed, clearly, but enhanced by a factor of $1/\epsilon$. So, the regime where the effect becomes in principle sizable is the one of eternal inflation, which is in fact characterized by $\alpha/\epsilon \gtrsim1$. 
\item \emph{Higher ``time" derivatives.}  As WdW features a field-space d'Alambert operator $\Box$, it obviously contains higher $\chiTwo-$derivatives. We argued in Sec.~\ref{highertime} that such terms are effectively $\alpha$-suppressed. However, also in this case one should notice that a compensating $1/\epsilon$ factor is present which make them in principle important during eternal inflation.   
\end{enumerate}

\subsection{Spatially flat gauge}
The change of variable in this case is given by $(\rho,\chiTwo)\rightarrow(\rho,\varphi)$ where $\rho$ plays the role of time and $\varphi=\chiTwo-\bar{\chiTwo}(\rho)$ describes the ``spatial" displacement from the classical solution. After applying the chain rules the derivatives transform according to 
 \begin{align}
 \partial_x &\rightarrow \partial_\varphi \\
 \partial_\rho &  \rightarrow \partial_\rho - \bar \chiTwo'(\rho) \partial_\varphi
 \end{align}
and at leading-order in  slow-roll eq. \eqref{eq:SCH_eq_slowroll} becomes
\begin{multline}\label{Scheq1_spat_flat}
    i\left[\left(1-\frac{\epsilon\eta}{4}(\rho+\varphi)^2\right)\partial_\rho\psi+\frac{\psi}{2}\mybox\! S + \left(\frac{3-\epsilon}{6}\eta\varphi+\frac{\epsilon\eta}{4}(\varphi^2+2\varphi\rho)\right)\partial_{\varphi}\psi \right]=\\-\frac{\alpha}{2
}e^{-(3-\epsilon)\rho+\frac{\epsilon\eta}{4}\rho^2}e^{\epsilon\varphi}\left[\left(-\frac{1}{6}\bar \chiTwo'^2+\frac{1}{2\epsilon}\right)\partial_\varphi^2\psi+\frac{1}{6}\bar \chiTwo''\partial_\varphi\psi\right].
\end{multline}

This is a clear example where the coordinate transformation has not managed to get rid of all spatial (i.e. $\varphi-$) derivative terms on the LHS of~\eqref{Scheq1_spat_flat}. The reason is that the $\varphi$ coordinate is not comoving in the sense described in Sec.~\ref{classflow} and we are in the situation depicted in Eq.~\eqref{matching2}.  

As explained in Sec.~\ref{apples} we need a phase-shift redefinition of the wavefunction of  the type 
\begin{equation}  
\tilde\psi = \psi \exp\left(-\frac{i f(\varphi, \rho)}{\alpha}\right).   
\end{equation}  
to get rid of this term. By inspection we find that  
\begin{equation}  
f(\varphi,\rho)\equiv 6 e^{-\left((-3 + \epsilon) \rho + \epsilon \varphi\right)}  
\left[
\frac{  
\eta \left(  
12 + \epsilon^2 \varphi (-2 + 6 \rho + 3 \varphi) + 2 \epsilon (-1 + 3 \rho + 6 \varphi)  
\right)  
}{  
12 (-3 + \epsilon) \epsilon  
}  
\right].  
\end{equation}  
When re-written for $\tilde \psi$, the left-hand side of Eq.~\eqref{Scheq1_spat_flat} no longer contains a space derivative (i.e., a derivative with respect to \(\varphi\)). However, the equation now acquires a mass term at order \(1/\alpha\), which matches precisely the mass term in the perturbative Schrödinger equation in the spatially flat gauge \eqref{eq:sch_sfg_perturbation1}. 

This result confirms that a change in the phase of the wave function eliminates the spatial derivative of the wave function at order ${\cal O}(\alpha^0)$ in the Schrödinger equation and introduces a mass term that precisely matches the expected mass term in the perturbative approach to order ${\cal O}(\alpha^{-1})$.

\section{Discussion and future plans} \label{conclusion}

Primordial inflation is a great example of a quantum gravity experiment conducted at energies lower than the Planck scale, i.e., within the regime of validity of general relativity. In this regime there are two almost equivalent---but not obviously identical---approaches to quantum gravity. In standard perturbation theory one writes a generic metric as a classical background solution plus perturbations and expands the Lagrangian of the system at the desired order in these perturbations. The latter can be described quantum mechanically by a wavefunction $\psi_P$. Or, one can quantize the system as it is, without reference to a background and write the WdW equation for the nonpertubative wavefunction $\Psi$.
 
In this paper we have tried to highlight the relation between $\Psi$ and $\psi_P$ (see eq.~\eqref{relationPW} and the examples in Secs.~\ref{scaling} and~\ref{slowroll} ). We have done this a little pedantically and ``by hands". It would be nice to develop an \emph{effective field theory} approach to include the WdW corrections in the perturbative framework in a more systematic way. To this end, it could perhaps be helpful to treat the wavefunctions as fields, write their Lagrangians and use some more standard QFT methods of approximation---instead of dealing with partial differential equations all the time. 
 
One other important spin-off of this work will be to include cosmic time as an independent degree of freedom in this quantum cosmology framework. In this paper we have used a classical solution, or a congruence of classical solutions, to translate the value of a field (say, $q^0$)  used as time into the proper time $t$ of the cosmic comoving observers. This relation is classical and one-to-one, but it should instead contain some uncertainty, representing the fact that the geodesic of the cosmic observer can be embedded in a given spacetime with a superposition of initial values of their clocks (see e.g.~\cite{Witten:2023qsv,Nitti:2024iyj}). We plan to turn to this problem next.

\section*{Acknowledgments} We acknowledge useful conversations and exchanges with Niayesh Afshordi, Thomas Colas, Angelo Esposito, Julien Grain, Lam Hui, Oliver Janssen, Matthew Johnson,    Jean-Luc Lehners,  Mehrdad Mirbabayi,  Alberto Nicolis, Alessandro Podo,   Sergey Sibiryakov, Alexander Taskov, Andrew Tolley and Vincent Vennin. This work received support from the French government under the France 2030 investment plan, as part of the Initiative d'Excellence d'Aix-Marseille Universit\'e - A*MIDEX (AMX-19-IET-012). This work  is also supported by the Programme National GRAM of CNRS/INSU with INP and IN2P3 co-funded by CNES.

\appendix

\section{More about the Hamilton Jacobi equation} \label{sec_HJ}

\subsection{The Klein Gordon example}
\label{sec_KG}

Sidney Coleman  famously stated that the career of a theoretical physicist consists of treating the harmonic oscillator in ever-increasing levels of abstraction. For those who want to waste their time on quantum gravity an equally important paradigm is certainly the Klein Gordon (KG) equation, which is a simple example of~\eqref{wdw} with $\alpha =1$, $g_{\mu \nu} = \eta_{\alpha \beta}$ and  $U(q) = m^2/2$. 
\begin{equation} \label{fullkg}
\left(\eta^{\alpha \beta}\partial_\alpha \partial_\beta   - m^2\right)\Psi = 0\, .
\end{equation}
The classical solutions are straight lines in Minkowski space. For example, those that go through the origin can be parameterized by their momentum $\vec p$,
\begin{equation} \label{KG_traj}
\bar q^0 = E \, t, \qquad \vec {\bar q} = \vec p \, t, 
\end{equation}
with $E = \sqrt{m^2 + \vec p^{\; 2}}$ from the constraint equation~\eqref{constraineq}. The Hamilton Jacobi equation reads 
\begin{equation}
(\partial_0 S)^2 - (\partial_i S)^2 - m^2 = 0\, .
\end{equation}

One can always  find a solution $S$ such that 
\begin{equation}
\nabla^\mu S = \dot {\bar q}^{\, \mu}\, 
\end{equation}
on the trajectory. 
This is apparent from~\eqref{constraineq} and~\eqref{HJ}. We call a solution $S$ \emph{associated} to the trajectory ${\bar q}^{\, \mu}$ if the above applies. However, the associated solution is far from unique. 
For example,  the plane wave
\begin{equation} \label{flat_wave}
S_{plane}(q) = - E\, q^0 + \vec p \cdot \vec q\, ,
\end{equation}
is associated to~\eqref{KG_traj}.
But another legitimate choice is the ``hyperboloid" solution 
\begin{equation} \label{hyp_wave}
S_{hyp}(q) = m \sqrt{-q_\mu q^\mu}\, .
\end{equation}

\begin{figure}[h]
%\vspace{-1cm}
\begin{center}
     \includegraphics[width=7cm]{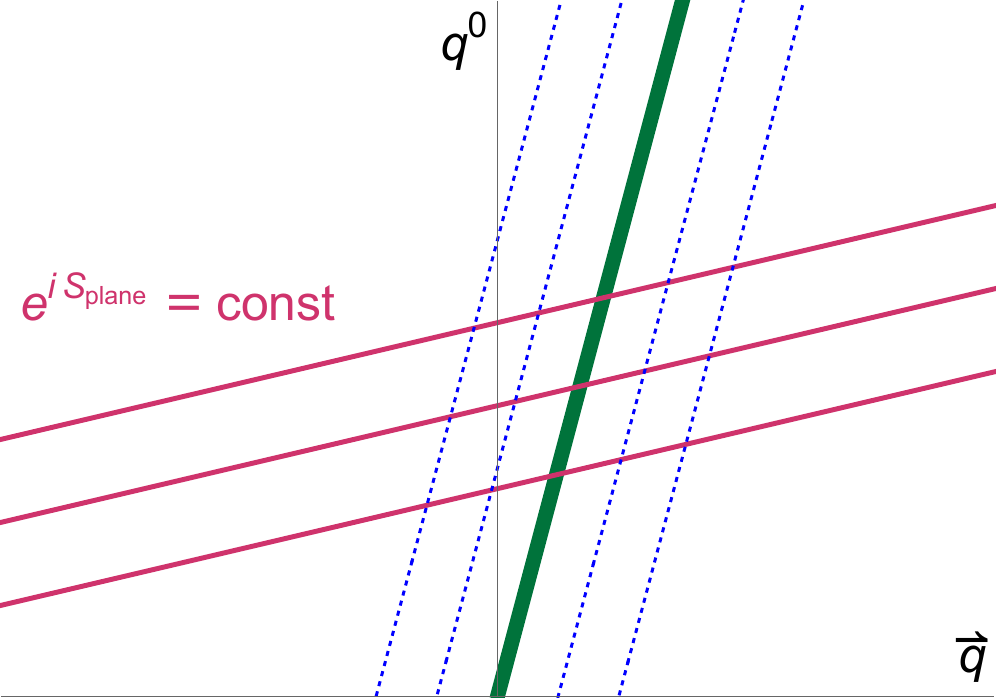}\ \ \ \ \ 
    \includegraphics[width=7cm]{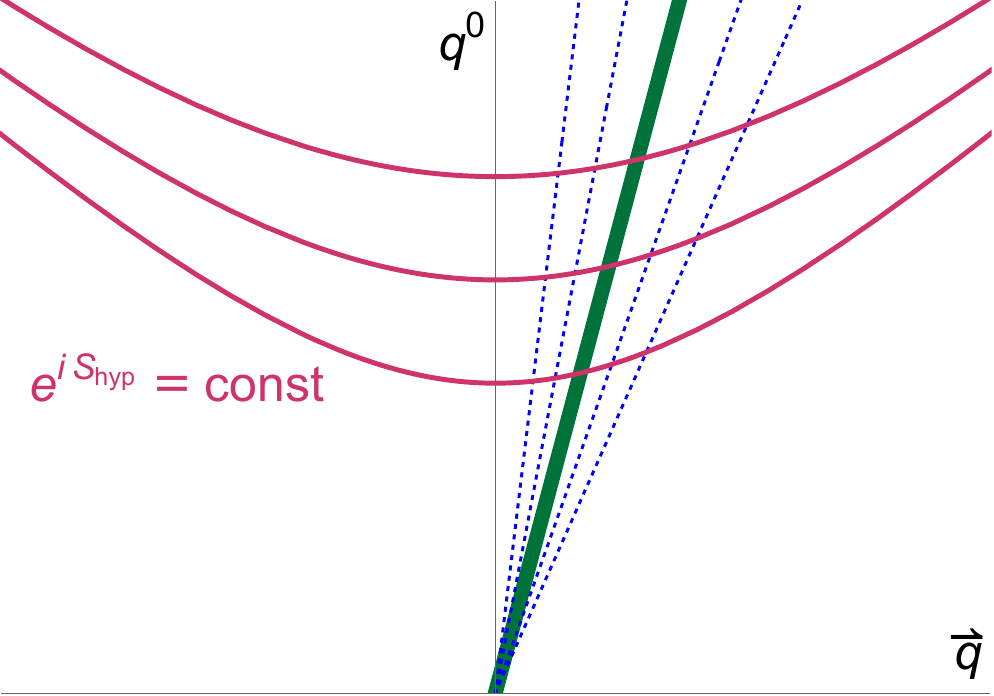}
  \end{center}\vspace{-.4cm}
  \captionsetup{width=.95\linewidth}
\caption{\small Two examples of solutions of the HJ equation associated with a given classical trajectory (think, green). Dotted in blue, some neighboring trajectories for each case. In the notation introduced in Sec.~\ref{classflow}, such curves are examples of $\zeta^i = const.$ curves because they correspond to potential solutions of the classical equations of motion. } \label{fig-1}
  \end{figure}
  
Different choices of $S$ can be seen as different congruences of classical trajectories to which  $\bar q^{ \, \mu}(t)$ can belong. For example, 
the nearby classical trajectories that are orthogonal to the flat wavefront~\eqref{flat_wave} are all parallel to each other while those orthogonal to the hyperboloid share a common point at $q^\mu=0$ (Fig.~\ref{fig-1}). 
For many purposes it is most convenient to choose (at least, locally-) flat wavefronts for $S$. However this is generally not possible beyond this KG example, as we show below.

\subsection{Use of Fermi coordinates} \label{fermi}

The above example shows that $S$ is all but unique. While $\nabla_\mu S$ is fixed by $\bar q^{ \, \mu}(t)$, $\nabla^2 S$ is free to vary. In other words, the wavefront $S$ can be a curved surface in field space.  
Since we are interested in approximating $\Psi$ close to the classical trajectory we can try to be pragmatic and choose $S$ in such a way that $\nabla^2 S =0$ on $\bar q^{ \, \mu}(t)$. However, this can be done in the KG case but not in general. In order to see this, let us build a Fermi coordinate system $x^\mu$ around the classical trajectory.\footnote{In the highly symmetric example of the KG equation going to Fermi coordinates merely corresponds to boosting the classical solution~\eqref{KG_traj} into  $\bar q^0 = m t, \vec {\bar q} = 0$.}

First, notice that $\bar q^{ \, \mu}(t)$ is accelerated in virtue of the potential term in~\eqref{fieldeq}. Also, by~\eqref{constraineq}, the proper time $\tau$ along the curve is related with the parameter $t$ through $d\tau = \sqrt{2U} dt$. The proper acceleration then reads
\begin{equation} \label{acceleration}
a^\mu \, = \, \frac{D}{\partial \tau}\frac{\partial}{\partial \tau}\, \bar q^{ \, \mu} \, = \, -\frac{1}{2 U} h^{\mu \nu} \partial_\nu U,
\end{equation}
where $h$ is the projector orthogonal to $\bar q^{ \, \mu} $.

The metric components in Fermi coordinates (by construction, $x^0 =\tau$) read
\begin{align}\label{00}
 g^{00} &= -1 + 2 a_i (\tau) x^i +  M_{ij}(\tau) \, x^i x^j + {\cal O}(x)^3 \, , \\
 g^{0i} &= {\cal O}(x)^2 \, ,\\
 g^{ij} &= \delta_{ij} + {\cal O}(x)^2\, , \label{11}
 \end{align}
where $M_{ij}$ is a combination of components of the  Riemann tensor and of the acceleration (see e.g.~\cite{Nesterov:1999ix}). Notice that, by~\eqref{acceleration}, $a_0=0$ in these coordinates. A similar ``tubular" expansion can be done for $S$, 
\begin{equation} \label{S}
S(x^\mu) = - \int^\tau \!d\tau ' m(\tau') + \frac12 S_{ij}(\tau)\, x^i x^j + {\cal O}(x)^{ 3}\, .
\end{equation}
Terms linear in $x^i$ are excluded because the gradient of $S$ at $x^i = 0$ is parallel to the curve. Making the wavefront of $S$ locally flat around the classical trajectory would correspond to setting $S_{ij}=0$. We see in the following that this condition, in general, cannot be preserved in time.  The potential $U$ can also be expanded around $x^i = 0$. 
 
By sticking these ingredients inside the Hamilton Jacobi equation~\eqref{HJ} we obtain
\begin{equation}
 -m^2 + m S'_{ij}x^i x^j + 2 m^2 a_i x^i + m^2  M_{ij} x^i x^j + S_{ki} S_{kj}x^i x^j 
+ 2 U + 2 \partial_i U x^i + \partial_i \partial_j U x^i x^j  \, =\,  {\cal O}(x)^3\, .
\end{equation}
By equating the terms at $x^i=0$ we get
\begin{equation}
m^2 (\tau)= 2 U(\tau)\, ,
\end{equation}
which nicely cancels the terms linear in $x^i$ using~\eqref{acceleration}. The terms quadratic in $x^i$ give
\begin{equation}
\setlength{\fboxsep}{2\fboxsep}\boxed{m S'_{ij} + S_{ki} S_{kj} = - m^2 M_{ij} -  \partial_i \partial_j U \, .}
\end{equation}
Not surprisingly, this is reminiscent of the Raychaudhuri equation, governing the expansion of a congruence of curves. We see from the above that both the potential $U$ and the Riemann curvature in field space act as sources for $S_{ij}$. Meaning, in general, that the wavefront $S= const.$ cannot remain flat all along the classical solution $\bar q^{ \, \mu}(t)$.

\renewcommand{\baselinestretch}{1}\small
\bibliographystyle{ourbst} %REMOVED DUE TO BIBLATEX
%\bibliography{replicaBib}
\bibliography{references} %REMOVED DUE TO BIBLATEX

\end{document}